\documentclass[aps,prd,preprintnumbers,nofootinbib]{revtex4}
\usepackage[dvipdfmx]{graphicx}
\usepackage{bm,latexsym,amsmath,amsthm,amssymb,amsfonts,color}

\usepackage[normalem]{ulem}

\begin{document}

\title{\uline{}Loosely trapped surface for slowly rotating black hole}

\author{${}^{1,2}$Keisuke Izumi, ${}^{1,2}$Tetsuya Shiromizu, ${}^1$Daisuke Yoshida, ${}^{3}$Yoshimune Tomikawa and ${}^{4,5}$Hirotaka Yoshino}

\affiliation{${}^1$Department of Mathematics, Nagoya University, Nagoya 464-8602, Japan}
\affiliation{${}^2$Kobayashi-Maskawa Institute, Nagoya University, Nagoya 464-8602, Japan}
\affiliation{${}^3$Division of Science, School of Science and Engineering, Tokyo Denki University, Saitama 350-0394, Japan}
\affiliation{${}^4$Department of Physics, Osaka Metropolitan University, Osaka 558-8585, Japan}
\affiliation{${}^5$Nambu Yoichiro Institute for Theoretical and Experimental Physics (NITEP), 
Osaka Metropolitan University, Osaka 558-8585, Japan}
%
%
\begin{abstract}
We construct the marginal loosely trapped surface (marginal LTS) for the Kerr spacetime with a small Kerr parameter 
perturbatively, where the LTS condition is saturated. 
An LTS is a surface that specifies the strong gravity region, which is a generalization of the photon sphere in the Schwarzschild spacetime. 
It turns out that there are an infinite number of marginal LTSs.
At the leading order of the small Kerr parameter, all of the marginal LTSs have the same area. 
However, one can see that the maximal marginal LTS among them is 
uniquely determined at the higher order. 
\end{abstract}

\maketitle

%
%
\section{introduction} 

A loosely trapped surface (LTS) has been introduced by four of the current authors to describe the strong gravity region 
outside black holes~\cite{Shiromizu2017}. Under certain reasonable conditions, one can show that the area of an LTS satisfies 
the Penrose-like inequality. Since an LTS is defined by an inequality controlling the strength of gravity, a large number of LTSs can 
exist in a spacetime. 
The marginal LTS is defined as an LTS where the inequality is saturated. 
In the Schwarzschild black hole, the photon sphere, which is composed of the closed circular orbits of photons, is 
the marginal LTS. In addition, a surface of constant area radius inside the photon sphere is an LTS, and the area bound theorem 
in Ref.~\cite{Shiromizu2017} tells us that the photon sphere is the unique LTS with the largest area among all LTSs, which we call 
the maximum LTS. For the Kerr black hole, the candidate for an LTS has been also presented numerically~\cite{Yoshino2017a}. 
However, there has been no systematic study to find the marginal and/or maximum LTS even in the Kerr black hole.

In this paper, we tackle the problem of finding the marginal and/or maximum LTS perturbatively for 
the Kerr spacetime with a small Kerr parameter $a$. 
In the current setup, where the deviation from the Schwarzschild spacetime is tiny, the locus of the marginal LTS can be supposed to be 
$r=3M(1+\delta)$ with $\delta=\mathcal{O}(a^2)$, $r$ being the area radius and $M$ the Arnowitt-Deser-Misner (ADM) mass. 
Naively, the marginal LTS is expected to be the one with the maximum area. 
Our purpose is to determine $\delta$ for the marginal LTS. 
However, up to $\mathcal{O}(a^2)$, we will see that there are an infinite number of the marginal LTSs. 
Furthermore, all of these marginal LTSs have the same area at $\mathcal{O}(a^2)$. 
In the case of the Schwarzschild spacetime, thanks to the area bound theorem, the maximum marginal LTS is unique. 
This property is expected to hold, meaning that at the higher order of $a$, a unique maximum marginal LTS should be found.
Therefore, we shall compute the area of the marginal LTS up to $\mathcal{O}(a^4)$, and find the maximum marginal LTS uniquely. 

The remaining part of this paper is organized as follows. In Sec. II, we briefly review the LTS and 
the setup for the Kerr black hole, and we will introduce the marginal LTS. In Sec. III, 
we shall determine the locus of the marginal LTS for the Kerr black hole up to $\mathcal{O}(a^2)$. In Sec. III, we compute the area of 
the marginal LTS up to $\mathcal{O}(a^4)$ and then show that the deformation $\delta$ is uniquely determined up to $\mathcal{O}(a^2)$ for the 
maximal marginal LTS. In Sec. IV, we give the summary and discussion.  

%
%
\section{loosely trapped surface and setup}

In a spacelike hypersurface $\it \Sigma$ with an induced metric $q_{ab}$, a loosely trapped surface (LTS), $S$, is defined 
in Ref. \cite{Shiromizu2017} as a compact 2-surface such that the mean curvature $k = D_{a} r^{a}$ on $S$ is positive, $k|_{S} > 0$, 
and the derivative of $k$ along the outward direction evaluated on $S$ is non-negative, $r^{a} D_{a} k|_{S} \geq 0$, for some 
foliation including $S$ as a leaf. Here $r^{a}$ is the outward unit normal of each leaf and $D_{a}$ represents the covariant 
derivative with respect to the induced metric $q_{ab}$. 
Interestingly, for the area of LTSs, $A_{\rm LTS}$, in the asymptotically flat spacelike hypersurface $\it \Sigma$ with 
non-negative scalar curvature, one can show that 
\begin{eqnarray}
A_{\rm LTS} \leq 4\pi (3M)^2
\end{eqnarray}
holds, where $M$ is the ADM mass. The equality holds if and only if the induced metric of $\it \Sigma$, $q_{ab}$, is 
isometric to that of the static time-symmetric slice of the Schwarzschild spacetime. We also 
define the marginal LTS as an LTS such 
that the derivative of $k$ along the outward direction vanishes%
\footnote{
In general spaces, even if LTSs can be found, there is no guarantee that the marginal LTS exists. 
However, in axisymmetric smooth spaces, if LTSs exist, a marginal LTS is expected to be found due to the symmetry.
}, 
that is,  $r^{a}D_{a} k = 0$. 

We consider the marginal LTS for the Kerr black hole with the slow rotation. The metric is 
\begin{eqnarray}
g=g_{\mu\nu}dx^\mu dx^\nu=-\frac{\Delta \Sigma}{{\cal A}}  dt^2+\gamma^2(d\phi-\omega dt)^2+\varphi^2dr^2+\psi^2 d\theta^2,
\end{eqnarray}
where 
\begin{eqnarray}
\gamma^2:=\frac{{\cal A}\sin^2 \theta}{\Sigma}, \quad  \omega:=\frac{2Mar}{{\cal A}}, \quad  \varphi^2:=\frac{\Sigma}{\Delta},\quad  \psi^2:=\Sigma
\end{eqnarray}
and
\begin{eqnarray}
\Delta:=r^2+a^2-2Mr, \quad  \Sigma:=r^2+a^2 \cos^2 \theta, \quad  {\cal A}:=(r^2+a^2)^2-\Delta a^2 \sin^2 \theta.
\end{eqnarray}

In general, one can suppose that the marginal LTS $S$ is located at 
\begin{eqnarray}
r=R(x^A) \label{r=R}
\end{eqnarray}
in the $t=$const. hypersurfaces $\it \Sigma$ with the induced metric 
\begin{eqnarray}
q=q_{ij}(x^k)dx^idx^j=\varphi^2 dr^2+q_{AB}dx^A x^B=\varphi^2 dr^2+\psi^2d\theta^2+\gamma^2 d\phi^2,
\end{eqnarray}
where $x^{A}$ represents the coordinates on $S$: $\theta$ and $\phi$. 
The induced metric of $S$ is given by  
\begin{eqnarray}
h_{AB}(x^C)dx^Adx^B=(\bar h_{AB} +\bar \varphi^2 R_{,A}R_{,B}) dx^Adx^B,
\label{meth}
\end{eqnarray}
where $\bar h_{AB}(x^C)=q_{AB}(r=R(x^C),x^D)$. The unit normal vector to the marginal LTS is 
\begin{eqnarray}
r_a|_S= \left. \chi (dr-R_{,A}d x^A )_a \right|_S
\end{eqnarray}
and
\begin{eqnarray}
r^a|_S=\chi \left. \Bigl( \frac{1}{\bar \varphi^2} \partial_r- q^{AB} R_{,B} \partial_A \Bigr)^a\right|_S , \label{ronS}
\end{eqnarray}
where $\bar \varphi = \varphi (r=R(x^A),x^B)$ and $\chi$ is determined through the normalization as 
\begin{eqnarray}
\chi =\dfrac{\bar \varphi}{{\sqrt {1+\bar \varphi^2 \bar h^{AB} R_{,A}R_{,B}}}}. 
\end{eqnarray}

To calculate the first derivative of $k$, we have to introduce a foliation near 
the marginal LTS. In general, the unit normal vector to each leaf of 
the foliation near the LTS is written as 
\begin{eqnarray}
r^a=\frac{1}{\alpha} (\partial_r-\beta^A \partial_A)^a
\end{eqnarray}
where, in the current setup, the function
$\alpha$ is given by 
\begin{eqnarray}
\alpha={\sqrt {\varphi^2+q_{AB}\beta^A \beta^B}}
\end{eqnarray}
and, for the vector $\beta^A$,
 the comparison to Eq. (\ref{ronS}) implies us 
\begin{eqnarray}
\beta^A|_S=\left. \varphi^2 q^{AB}R_{,B} \right|_S . \label{betaonS}
\end{eqnarray}
In the next section, we will assume that the marginal LTS is located at 
$r=R=3M+\mathcal{O}(a^2)$. In this case, Eq. (\ref{betaonS}) implies us 
$\beta^A=\mathcal{O}(a^2)$, which gives an estimation of $\alpha$ as $\alpha= \varphi +\mathcal{O}(a^4)$. 
The vector $\beta^A$ depends on the foliation, the choice of which is free except on $S$. 
Therefore, $\beta^A$ can be arbitrary except on $S$ (but it should be smooth due to the smooth foliation).

\section{marginal LTS: up to $\mathcal{O}(a^2)$}

In this section, for a small Kerr parameter $a$, we derive the first derivative of the mean curvature $k$ of $S$ along 
the outward direction up to $\mathcal{O}(a^2)$ and find the marginal LTS, satisfying $r^aD_a k|_S=0$. 

The area bound theorem proven in Ref.~\cite{Shiromizu2017} shows that the photon sphere ($r = 3M$) is the maximum LTS 
and hence the equality in the definition of the LTS holds; that is, the photon sphere is the marginal LTS. 
Since the spacetime smoothly approaches the Schwarzschild spacetime as the Kerr parameter $a$ is turned off, 
the marginal LTS of the Kerr spacetime with a small Kerr parameter $a$ is expected to exist around $r=3M$. 
Since the corrections to the induced metric of $t=$const. hypersurface of the Kerr spacetime from the Schwarzschild spacetime (see Eq.~\eqref{meth}) 
are $\mathcal{O}(a^2)$, 
the position of the marginal LTS is expected to deviate by $\mathcal{O}(a^2)$ from $r=3M$.
Hence, by introducing a function $\delta_{1}(x^{A})$ of $\mathcal{O}(a^2)$, 
we can suppose $R=3M \Bigl(1+\delta_1(x^A) + \mathcal{O}(a^{4})\Bigr)$ for $R(x^A)$ in Eq. (\ref{r=R}), with which $k$ is computed as 
\begin{eqnarray}
k=D_ar^a 
=  \frac{1}{2\varphi}\frac{{\cal A}_{,r}}{{\cal A}}-\frac{1}{\varphi_0}{\cal D}_A \beta^A_1 +\mathcal{O}(a^4),
\end{eqnarray}
where ${\cal D}_A$ is the covariant derivatives with respect to the induced metric of $S^2$, $\varphi_0$ is defined as $\varphi_0:=\displaystyle \lim_{a \to 0}\varphi$, and we used $\alpha=\varphi+\mathcal{O}(a^4)$ and $\beta^A = \beta_{1}^{A} + \mathcal{O}(a^4)$ with $\beta_{1}^{A} = \mathcal{O}(a^2)$. 

Taking the normal derivative of $k$ to $S$, we have 
\begin{eqnarray}
\varphi r^aD_a k|_S =   -\frac{2}{3{\sqrt 3}M^2}\delta_1 +\frac{1}{81{\sqrt 3}M^2}(-7+\cos^2 \theta)\Bigl( \frac{a}{M} \Bigr)^2
-\frac{1}{3{\sqrt 3}M^2}{\cal D}^2\delta_1 -\frac{1}{{\sqrt 3}}{\cal D}_A \bar{\beta}^A_{1,r} +\mathcal{O}(a^4)=0
\label{eq:main}
\end{eqnarray}
on $S$, where ${\cal D}^2:= \sigma^{AB} {\cal D}_A {\cal D}_B$,  $\sigma_{AB}$ is the metric of the 2-dimensional unit round sphere $S^2$, 
$\bar \beta^A_{1,r}$ is defined by $\bar \beta^A_{1,r}:=\beta^A_{1,r}|_S$ and 
we used 
\begin{eqnarray}
\beta^A|_S=\beta^A_{1}|_S+\mathcal{O}(a^4)=\sigma^{AB} \delta_{1,B}/M+\mathcal{O}(a^4).
\end{eqnarray}

In general, $\delta_1$ and $\bar \beta^A_{1,r}$ on $S^2$ can be expanded by the spherical harmonic function $Y_{\ell m} =Y_{\ell m} (x^A)$ as%
\footnote{Since $\delta_1$ and the components of $\beta^A_{,r}$ are real, 
\begin{eqnarray}
a_{\ell -m} = a_{\ell m}^*, \hspace{10mm}
b_{\ell -m} = b_{\ell m}^*, \hspace{10mm}
c_{\ell -m} = c_{\ell m}^*, 
\end{eqnarray}
are satisfied.}
\begin{eqnarray}
\delta_1=\sum_{\ell,m}a_{\ell m} Y_{\ell m}
\end{eqnarray}
and
\begin{eqnarray}
\bar{\beta}^A_{1,r} =\frac{1}{M^2} \sum_{\ell,m} \Bigl( b_{\ell m} {\cal D}^A Y_{\ell m}+c_{\ell m} \epsilon^{AB}{\cal D}_B  Y_{\ell m} \Bigr),
\end{eqnarray}
where $\epsilon^{AB}$ is the Levi-Civita tensor on $S^2$. 
Note that we can set $b_{00}=0,~c_{00}=0$ without loss of generality because they do not contribute to $\bar \beta^A_{1,r}$ 
due to the constantness of $Y_{00}$. 
Since $\beta^A$ can be arbitrary except on $S$, the normal derivative of $\beta_A$ on $S$, that is, $\bar{\beta}^A_{,r}$, can also be arbitrary. 
Therefore, $b_{\ell m}$ and $c_{\ell m}$ are parameters that we can choose arbitrarily.
After a short calculation, we show that the condition for the marginal LTS becomes 
\begin{eqnarray}
0&=& \varphi r^aD_a k|_S\nonumber \\
 & = &  
-\frac{2}{3{\sqrt 3}} \frac{1}{M^2}\Biggl( \frac{20 {\sqrt \pi}}{81}\Bigl(\frac{a}{M}\Bigr)^2+a_{00}  \Biggr) Y_{00}
+\frac{2}{3{\sqrt 3}} \frac{1}{M^2}\Biggl( \frac{2 }{81}{\sqrt {\frac{\pi}{5}}} \Bigl(\frac{a}{M}\Bigr)^2 +2a_{20}+9b_{20} \Biggr) Y_{20} \nonumber \\
& & 
+\frac{1}{3{\sqrt 3}M^2} \sum_{(\ell,m) \neq (0,0) ,(0, 2)} \Biggl[ (\ell+2)(\ell-1)a_{\ell m}+3 \ell (\ell+1)b_{\ell m}  
\Biggr] Y_{\ell m} + \mathcal{O}(a^4). 
\end{eqnarray}
This implies us 
\begin{eqnarray}
a_{00}= -\frac{20 {\sqrt \pi}}{81} \Bigl( \frac{a}{M} \Bigr)^2, \label{a00}
\end{eqnarray}
\begin{eqnarray}
b_{1m}=0, \label{b1m}
\end{eqnarray}
\begin{eqnarray}
a_{20}= - \frac{1}{81}{\sqrt {\frac{\pi}{5}}} \Bigl( \frac{a}{M} \Bigr)^2 -\frac92 b_{20}, \label{a20}
\end{eqnarray}
\begin{eqnarray}
(\ell+2)(\ell-1)a_{\ell m}+3 \ell (\ell+1)b_{\ell m}=0  \hspace{10mm} \mbox{for $(\ell,m) \neq (0, 0),  (2,0)$}, 
\label{almblm}
\end{eqnarray}
and $a_{1m}$ can be any value. 
Except for $a_{00}$, $a_{\ell m}$'s can take any value due to the arbitrariness of $b_{\ell m}$. 
This means that, up to $\mathcal{O}(a^2)$, any deformation except those including the homogeneous 
component keeps the surface as a marginal LTS, 
namely, satisfying $\varphi r^aD_a k|_S =0$ by adjusting the foliation near $S$. 

It is easy to compute ${\sqrt {\det (h_{AB})}}$ as 
\begin{eqnarray}
{\sqrt {\det (h_{AB})}}|_S = \psi \gamma|_S =  (3M)^2 \sqrt{\det (\sigma_{AB})} \Biggl[ 1 + 2\delta_1+ \frac{1}{9} \Bigl(\frac{a}{M} \Bigr)^2 \left(1-\frac{1}{6} \sin^2\theta\right)\Biggr] + \mathcal{O}(a^4). \label{deth}
\end{eqnarray}
Then, noting the existence of the contribution from the monopole component for $\delta_1$, the area of $S$ is written as 
\begin{eqnarray}
A_S=4\pi (3M)^2 \Biggl[ 1-\frac{4}{27}\Bigl(\frac{a}{M}\Bigr)^2 \Biggr] +\mathcal{O}(a^4).
\end{eqnarray}
Obviously, it does not depend on the higher multipole moments of $\delta_1$, that is, the choice of a LTS. 

In summary, up to $\mathcal{O}(a^2)$, we have an infinite number of the marginal LTSs with the same area.

\section{marginal LTS with maximum area} 

Because of the arbitrary choice of $b_{\ell m}$, we saw in the previous 
section that almost any deformed surface is a marginal LTS. 
Here, we try to find the marginal LTS with the maximum area among them ({\it i.e.,} the maximal marginal LTS). 
In the case of the Schwarzschild spacetime, the area bound theorem proven in Ref.~\cite{Shiromizu2017} shows that the maximum area is 
achieved for the surface with $r=3M$. 
Since the introduction of the Kerr parameter $a$ continuously deforms the metric from the Schwarzschild spacetime, 
we expect that the configuration space consisting of marginal LTS's has qualitatively the same, that is, 
the extremal LTS around $r=3M$ must be the maximum. 
The result in the previous section implies that, even in the Schwarzschild spacetime, any $\mathcal{O}(\delta_1)$ deformation of the surface from $r=3M$ does not change 
the area up to $\mathcal{O}(\delta_1)$. 
However, the surface at $r=3M$ should be the maximum marginal LTS due to the area bound theorem, 
which is expected to be confirmed in the higher order analysis. 
Therefore, we proceed to the analysis at the order of $a^4$. 

Again, we suppose that the marginal LTS is located at $r=R(x^A)$, where 
\begin{eqnarray}
R=3M(1+\delta)=3M\left(1+\delta_1+\delta_2 + \mathcal{O}(a^6)\right)
\end{eqnarray}
with $\delta_1=\mathcal{O}(a^2)$ and $\delta_2=\mathcal{O}(a^4)$. The function $\alpha$ and the vector $\beta^{A}$ are expanded as 
\begin{eqnarray}
\alpha=\varphi \Bigl(1+\frac{1}{2\varphi^2}q_{AB}\beta^A \beta^B  \Bigr) +\mathcal{O}(a^6)
\end{eqnarray}
and
\begin{eqnarray}
\beta^A=\beta_1^A+\beta_2^A +\mathcal{O}(a^6),
\end{eqnarray}
where $\beta^A_1=\mathcal{O}(a^2)$ and $\beta^A_2=\mathcal{O}(a^4)$. 

Let us derive the area of the marginal LTS. 
To do so, we shall consult the marginal condition $\varphi r^aD_ak|_S=0$ up to $\mathcal{O}(a^4)$, because the area expression 
includes a linear term proportional to $\delta_2$, which will be determined by the marginal condition at 
the order of $\mathcal{O}(a^4)$. As seen soon, however, note that the calculation for the area does not require the explicit form of $\beta^A_2$. 

We first compute the mean curvature for each leaf  
\begin{eqnarray}
k & = & \frac{2}{r}{\sqrt {1-\frac{2M}{r}}} \Biggl[1+f_1 \Bigl(\frac{a}{r} \Bigr)^2-\frac{r}{2}{\cal D}_A \beta^A
+f_2 \Bigl(\frac{a}{r} \Bigr)^4-\frac{r^2}{2}\Bigl(2-\frac{3M}{r} \Bigr)\sigma_{AB}\beta^A \beta^B \nonumber \\
& & +\frac{r}{4}\Bigl( \cos^2 \theta -\frac{1}{1-\frac{2M}{r}}\Bigr){\cal D}_A \beta^A\Bigl(\frac{a}{r} \Bigr)^2
-\frac{r^3}{2}\Bigl(1-\frac{2M}{r} \Bigr) \sigma_{AB}\beta^A_{,r}\beta^B
+\frac{r}{4}\Bigl(1-\frac{2M}{r} \Bigr) \beta^A {\cal D}_A \sin^2 \theta  \Bigl(\frac{a}{r} \Bigr)^2
 \Biggr] +\mathcal{O}(a^6). 
\end{eqnarray}
Here $f_1$ and $f_2$ are 
\begin{eqnarray}
f_1:=-1-\frac{1}{2}\cos^2 \theta+\frac{1}{2}(1-3x)\sin^2\theta +\frac{1}{2(1-2x)}
\end{eqnarray}
and
\begin{eqnarray}
f_2:=f_{20}+f_{22}\cos^2 \theta+f_{24}\cos^4 \theta,
\end{eqnarray}
with $x:=M/r$, where $f_{20}$, $f_{22}$ and $f_{24}$ are defined as
\begin{eqnarray}
f_{20}:=\frac{1}{8(1-2x)^2}(1+2x-28x^2-16x^3+96x^4),
\end{eqnarray}
\begin{eqnarray}
f_{22}:=-\frac{1}{4(1-2x)}(1-4x+30x^2-48x^3)
\end{eqnarray}
and
\begin{eqnarray}
f_{24}:=\frac{9}{8}-\frac{13}{4}x+3x^2.
\end{eqnarray}
After very long calculation, we have $(\varphi r^aD_a k|_S)$ at the order of $\mathcal{O}(a^4)$ as 
\begin{eqnarray}
(\varphi r^aD_a k|_S)^{(4)} & = &  -\frac{1}{3{\sqrt {3}}M^2}(2\delta_2+{\cal D}^2 \delta_2)
+\frac{8}{3{\sqrt {3}}M^2} \delta_1^2+\frac{7}{3{\sqrt {3}}M^2} \delta_1 {\cal D}^2 \delta_1
-\frac{5}{3{\sqrt {3}}M^2}({\cal D}\delta_1)^2 \nonumber \\
& & +\frac{1}{{\sqrt {3}}M^2} {\cal D}_A \delta_1 {\cal D}^A{\cal D}^2\delta_1
-\frac{10}{{\sqrt {3}}}{\cal D}_A \delta_1 \beta^A_{1,r}
-\frac{1}{{\sqrt {3}}} {\cal D}_A \bar \beta^A_{2,r} 
-\frac{1}{{\sqrt {3}}} \delta_1 {\cal D}_A  \bar \beta^A_{1,r} 
-{\sqrt {3}}M^2 \bar \beta^A_{1,r} \bar \beta_{1A,r} \nonumber \\
& & +\frac{1}{M^2}\frac{4}{81}{\sqrt {\frac{\pi}{3}}}\Bigl(29Y_{00}+\frac{1}{{\sqrt {5}}}Y_{20} \Bigr) 
 \delta_1\Bigl( \frac{a}{M}\Bigr)^2 
+\frac{1}{M^2}\frac{2}{243}{\sqrt {\frac{\pi}{3}}}\Bigl(71Y_{00}-\frac{23}{{\sqrt {5}}}Y_{20} \Bigr) 
{\cal D}^2\delta_1 \Bigl( \frac{a}{M}\Bigr)^2 \nonumber \\
& & -\frac{1}{M^2}\frac{10}{243}{\sqrt {\frac{\pi}{15}}}{\cal D}_A {\cal D}_B Y_{20}
{\cal D}^A{\cal D}^B \delta_1 \Bigl( \frac{a}{M}\Bigr)^2 
+\frac{2}{27}{\sqrt {\frac{\pi}{3}}}\Bigl(-4Y_{00}+\frac{1}{{\sqrt {5}}}Y_{20} \Bigr) 
{\cal D}_A \bar \beta^A_{1,r} \Bigl( \frac{a}{M}\Bigr)^2 \nonumber \\
& & +\frac{1}{M^2}\frac{20}{243}{\sqrt {\frac{\pi}{15}}}{\cal D}_A  Y_{20}
{\cal D}^A \delta_1 \Bigl( \frac{a}{M}\Bigr)^2 
 -\frac{2}{81}{\sqrt {\frac{\pi}{15}}}{\cal D}_A  Y_{20} \bar \beta^A_{1,r}\Bigl( \frac{a}{M}\Bigr)^2 \nonumber \\
& & +\frac{1}{M^2}\frac{4}{2187}{\sqrt {\frac{\pi}{3}}}\Bigl(\frac{183}{5}Y_{00}+\frac{64}{7{\sqrt {5}}}Y_{20}
-\frac{26{\sqrt {\pi}}}{105}Y_{40} \Bigr) \Bigl( \frac{a}{M}\Bigr)^4,
\end{eqnarray}
where $\bar \beta^A =\beta^A (r=R(x^B),x^C)$, and we used 
\begin{eqnarray}
\left({\cal D}_A \beta^A \right)|_S={\cal D}_A \bar \beta^A-3M \beta^A_{,r} {\cal D}_A \delta_1 +\mathcal{O}(a^6)
\label{DAAS}
\end{eqnarray}
and 
\begin{eqnarray}
\left({\cal D}_A \beta^A_{,r} \right)|_S={\cal D}_A \bar \beta^A_{,r}-3M \beta^A_{,rr} {\cal D}_A \delta_1 +\mathcal{O}(a^6). 
\end{eqnarray}
In more detail, 
substituting Eq.~\eqref{betaonS} into Eq.~\eqref{DAAS}, we can express the $\mathcal{O}(a^4)$ parts of ${\cal D}_A \beta^A|_S$ as
\begin{eqnarray}
\left({\cal D}_A \beta^A\right) |_S ^{(4)}& = & -\frac{1}{3M}{\cal D}^2 \delta_1 \Bigl( \frac{a}{M}\Bigr)^2
+\frac{1}{M}{\cal D}^2 \delta_2
-\frac{4}{M}(\delta_1{\cal D}^2\delta_1+({\cal D}\delta_1)^2)
-3M \beta^A_{1,r}{\cal D}_A \delta_1 \nonumber\\
& & +\frac{5}{54M} \Bigl( 
{\cal D}_A{\cal D}_B \cos^2 \theta    {\cal D}^A{\cal D}^B \delta_1  
+2(2\cos^2\theta-1){\cal D}^2 \delta_1-{\cal D}_A \cos^2 \theta {\cal D}^A \delta_1
\Bigr)  \Bigl( \frac{a}{M}\Bigr)^2,
\end{eqnarray}
where we used 
\begin{eqnarray}
{\cal D}_\phi {\cal D}_\phi \delta_1=-\frac{1}{2}{\cal D}_A{\cal D}_B \cos^2 \theta    {\cal D}^A{\cal D}^B \delta_1  
-(2\cos^2\theta-1){\cal D}^2 \delta_1.
\end{eqnarray}

Let us compute the area of the marginal LTS up to $\mathcal{O}(a^4)$. Since
\begin{eqnarray}
{\sqrt {\det(h_{AB})}}|_S & = & (3M)^2 {\sqrt {\det (\sigma_{AB})}} \Biggl[ 
1+2\delta_1+\frac{1}{9}\Bigl(1-\frac{1}{6}\sin^2 \theta \Bigr)\Bigl(\frac{a}{M} \Bigr)^2
+2\delta_2+\delta_1^2-\frac{1}{27}\delta_1 \sin^2 \theta \Bigl(\frac{a}{M} \Bigr)^2
 \nonumber \\
& & ~~ -\frac{1}{243}\Bigl(1+\frac{1}{24}\sin^2 \theta \Bigr)\sin^2 \theta \Bigl(\frac{a}{M} \Bigr)^4 
+\frac{3}{2}({\cal D}\delta_1)^2  \Biggr] +\mathcal{O}(a^6),
\end{eqnarray}
we have 
\begin{eqnarray}
A_S & = & \int_{S^2} {\sqrt {\det (h_{AB})}}|_S d\theta d\phi   \nonumber \\
& = & (3M)^2\Biggl[ 4\pi+ 4 {\sqrt {\pi}}a_{00}+\frac{32\pi}{81}\Bigl(\frac{a}{M} \Bigr)^2
-\frac{124}{10935}\Bigl( \frac{a}{M} \Bigr)^4-\frac{4{\sqrt {\pi}}}{81} \Bigl(a_{00}-\frac{1}{{\sqrt {5}}}a_{20} \Bigr)\Bigl(\frac{a}{M} \Bigr)^2
\nonumber \\
& & +\int_{S^2}\Bigl(2\delta_2+\delta_1^2+\frac{3}{2}({\cal D}\delta_1)^2  \Bigr)d\Omega
\Biggr] \nonumber \\
& = & (3M)^2 \Biggl[ 4\pi -\frac{16\pi}{27}\Bigl(\frac{a}{M} \Bigr)^2
-\frac{2\cdot 37607\pi}{3^6 \cdot 5 \cdot 193}\Bigl( \frac{a}{M} \Bigr)^4
-\frac{386}{3}(a_{20}-a_{20,{\rm max}})^2 \nonumber \\
& & -\sum_{\ell, m \neq (0,0), (2,0)} \frac{3\ell^6+9\ell^5+\frac{23}{2}\ell^4+8\ell^3+\frac{15}{2}\ell^2+5\ell+4 }{\ell (\ell+1)}
|a_{\ell m}|^2+9\ell (\ell+1) |c_{\ell m}|^2
\Biggr] +\mathcal{O}(a^6),
\end{eqnarray}
where 
\begin{eqnarray}
a_{20,{\rm max}}:=\frac{131}{81 \cdot 193} {\sqrt {\frac{\pi}{5}}}\Bigl(\frac{a}{M} \Bigr)^2.
\end{eqnarray}
In the above, we used the conditions (\ref{a00}), (\ref{b1m}), (\ref{a20}) and (\ref{almblm}), and 
\begin{eqnarray}
\int_{S^2} \Bigl(2 \delta_2+\delta_1^2+\frac{3}{2}({\cal D}\delta_1)^2  \Bigr)d\Omega
& = & \int_{S^2} \Biggl[ 
9\delta_1^2-\frac{21}{2}({\cal D}\delta_1)^2-3({\cal D}^2\delta_1)^2-27M^2{\cal D}_A \delta_1 \beta^A_{1,r}-9M^4
\beta^A_{1,r}\beta_{1A,r} \nonumber \\
& & +\frac{4{\sqrt {\pi}}}{27}\Bigl(29Y_{00}+\frac{1}{{\sqrt {5}}}Y_{20} \Bigr)\delta_1 \Bigl( \frac{a}{M}\Bigr)^2
+ \frac{2{\sqrt {\pi}}}{81}\Bigl(71Y_{00}-\frac{33}{{\sqrt {5}}}Y_{20} \Bigr){\cal D}^2\delta_1 \Bigl( \frac{a}{M}\Bigr)^2 \nonumber \\
& & -\frac{10}{81}{\sqrt {\frac{\pi}{5}}} {\cal D}_A{\cal D}_BY_{20} {\cal D}^A{\cal D}^B\delta_1 \Bigl( \frac{a}{M}\Bigr)^2
+\frac{8M^2{\sqrt {\pi}}}{9} \Bigl(-Y_{00}+\frac{1}{3{\sqrt {5}}}Y_{20} \Bigr)  {\cal D}_A \beta^A_{1,r} \Bigl( \frac{a}{M}\Bigr)^2\nonumber \\
& & +\frac{4{\sqrt {\pi}}}{729} \Bigl(\frac{183}{5}Y_{00}+\frac{64}{7{\sqrt {5}}}Y_{20} -\frac{26{\sqrt {\pi}}}{105}Y_{40}
\Bigr)  \Bigl( \frac{a}{M}\Bigr)^4
\Biggr]d\Omega,
\end{eqnarray}
which is derived from the integration of the marginal condition at the order of $\mathcal{O}(a^4)$, that is, 
$(\varphi r^aD_a k|_S)^{(4)}=0$.  

From the expression, one can see that the marginal LTS has the maximum area for the small deformation up to $\mathcal{O}(a^4)$ having 
the following parameters 
\begin{eqnarray}
a_{00}= -\frac{20 {\sqrt \pi}}{81} \Bigl(\frac{a}{M}\Bigr)^2,\quad a_{20}=a_{20,{\rm max}} 
\end{eqnarray}
and 
\begin{eqnarray}
a_{\ell m}=0,\quad  b_{\ell m}=0,\quad  c_{\ell m}=0  \hspace{10mm} \mbox{for $(\ell,m) \neq (0, 0), (2,0)$}.
\end{eqnarray}

In summary, the deformation $\delta$ and area of the maximal marginal LTS are 
\begin{eqnarray}
\delta=\Biggl( -\frac{20 {\sqrt \pi}}{81}Y_{00}+\frac{131}{81 \cdot 193}{\sqrt {\frac{\pi}{5}}}Y_{20} \Biggr) \Bigl(\frac{a}{M} \Bigr)^2
\end{eqnarray}
and
\begin{eqnarray}
A_S^{\rm max}  = 4\pi (3M)^2 \Biggl[ 1 -\frac{4}{27}\Bigl(\frac{a}{M} \Bigr)^2
-\frac{37607}{3^6 \cdot 10 \cdot 193}\Bigl( \frac{a}{M} \Bigr)^4 \Biggr] +\mathcal{O}(a^6).
\end{eqnarray}

\section{summary and discussion}

In this paper, we have derived the marginal LTS for a slowly rotating black hole and have found that there are 
an infinite number of the marginal LTSs. 
Then, we could confirm from the perturbative argument that the 
marginal LTS with the maximum area is unique.  We also comment on 
the Schwarzschild case in our analysis. 
The marginal condition for the LTS with the maximum area uniquely fixes the surface to be the photon sphere, that is, 
$a_{\ell m}=b_{\ell m}=c_{\ell m}=0$. 
This is consistent with the case of the equality in the area bound theorem proven in Ref.~\cite{Shiromizu2017}. 
We remind that the LTS for the Kerr black hole presented as Fig. 2 in Ref.~\cite{Yoshino2017a} is an $r=$const. surface 
in a $t=$const. slice and it is not marginal one. 

From the current observation, it is natural to conjecture that 
the maximal marginal LTS is unique at the non-linear level if it exists. At least, one 
can address this issue numerically. As an extension of the LTS to weak gravity region, 
the attractive gravity probe surface was proposed in Refs.~\cite{Izumi2021,Izumi2023}, and a similar analysis for a wider class of 
spacetimes is also of interest. These are left for future studies.

%
%
\begin{acknowledgments}
This work was initiated by discussions between T.S., Gary Gibbons and Harvey Reall at Cambridge. 
K. I., T. S., D. Y. and H. Y. are supported by Grant-Aid for Scientific Research from Ministry of Education, Science, 
Sports and Culture of Japan JSPS(No. JP21H05189). K. I. and T. S. are supported by 
are supported by Grant-Aid for Scientific Research from Ministry of Education, Science, 
Sports and Culture of Japan JSPS(JP21H05182). 
K. I. is also supported by JSPS Grant-in-Aid for Scientific Research (C) (JP24K07046). 
T. S. is also supported by JSPS Grants-in-Aid for 
Scientific Research (C) (JP21K03551), Fund for the Promotion of Joint International Research (JP23KK0048)
and Grant-in-Aid for Scientific Research (A) (JP24H00183). 
D.Y. is also supported by JSPS KAKENHI Grant No.~JP20K14469. 
H.Y is in part supported by JSPS KAKENHI Grant Numbers JP22H01220, and is partly supported by Osaka Central 
Advanced Mathematical Institute (MEXT Joint Usage/Research Center on Mathematics and Theoretical Physics JPMXP0619217849). 
\end{acknowledgments}





\end{document}